\documentclass[twocolumn,english,aps,letterpaper]{revtex4}
\usepackage{mathptmx}
\usepackage[T1]{fontenc}
\usepackage[latin9]{inputenc}
\usepackage{geometry}
\usepackage{amsmath}
\usepackage{color}
\usepackage{graphicx}
\usepackage{amssymb}
\usepackage{babel}

\begin{document}

\title{The Bose-Hubbard ground state: extended Bogoliubov and variational
methods compared with time-evolving block decimation }

\author{Ippei Danshita$\,^{1}$ and Pascal Naidon$\,^{2}$,}

\email{pascal@cat.phys.s.u-tokyo.ac.jp}

\affiliation{$\,^{1}$Department of Physics, Faculty of Science, Tokyo University
of Science, Shinjuku-ku, Tokyo 162-8601, Japan\\
$\,^{2}$ERATO Macroscopic Quantum Project, JST, Tokyo, 113-0033 Japan}

\begin{abstract}
We determine the ground-state properties of a gas of interacting bosonic
atoms in a one-dimensional optical lattice. The system is modelled
by the Bose-Hubbard Hamiltonian. We show how to apply the time-evolving
block decimation method to systems with periodic boundary conditions,
and employ it as a reference to find the ground state of the Bose-Hubbard
model. Results are compared with recently proposed approximate methods,
such as Hartree-Fock-Bogoliubov (HFB) theories generalised for strong
interactions and the variational Bijl-Dingle-Jastrow method. We find
that all HFB methods do not bring any improvement to the Bogoliubov
theory and therefore provide correct results only in the weakly-interacting
limit, where the system is deeply in the superfluid regime. On the
other hand, the variational Bijl-Dingle-Jastrow method is applicable
for much stronger interactions, but is essentially limited to the
superfluid regime as it reproduces the superfluid-Mott insulator transition
only qualitatively.
\end{abstract}
\maketitle


\section{Introduction}

Optical lattices loaded with cold atomic gases have provided a fertile
experimental testing ground for the study of fundamental phenomena
exhibited by quantum degenerate gases~\cite{rf:bloch,rf:morsch,rf:lewenstein}.
One of the greatest advantages of these systems is that microscopic
theories can be directly compared with experiments without the use
of any fitting parameters thanks to the diluteness of atomic gases
and the flexible and precise controllability of parameters. In particular,
the Bose-Hubbard model~\cite{rf:fisher,rf:jaksch} has been successful
in treating systems of cold bosonic atoms confined in optical lattices
and explaining their various intriguing physical properties observed
in experiments, such as the superfluid-Mott insulator transition~\cite{rf:greiner,rf:stoeferle,rf:ian1},
the superfluid critical velocities~\cite{rf:sarlo,rf:mun}, and
the quantum depletion of condensates~\cite{rf:xu,rf:ian2}.

The one-dimensional (1D) Bose-Hubbard model has the following form
\begin{eqnarray}
\hat{H}_{{\rm bh}} & = & -J\sum_{l=1}^{M}(\hat{a}_{l}^{\dagger}\hat{a}_{l+1}+{\rm h.c.})+\frac{U}{2}\sum_{l=1}^{M}\hat{n}_{l}(\hat{n}_{l}-1)\label{eq:BHH}\end{eqnarray}
where $\hat{a}_{l}$ ($\hat{a}_{l}^{\dagger}$) creates a boson at
the lowest level localised on the $l$-th site of a 1D lattice, and
$\hat{n}_{l}=\hat{a}_{l}^{\dagger}\hat{a}_{l}$ is the number operator.
$J$ is the hopping energy from one site to the nearest neighbor,
and $U>0$ is the onsite repulsion. When the total number of particles
$N$ is incommensurate with the number of sites $M$, the system is
in a superfluid state at zero temperature. When $N$ is commensurate
with $M$, the system is superfluid for small $U/J$ and becomes a
Mott insulator for large $U/J$.

In the superfluid phase, several approximations are available to treat
this Hamiltonian analytically. The most famous one is the Bogoliubov
approximation \cite{rf:bogoliubov}. However, it is limited to very
weak interactions, i.e. $U/(nJ)\ll1$, where $n=N/M$ is the filling
factor. It would be desirable to extend this approximation to stronger
interactions. In the context of magnetically-trapped ultracold atomic
gases, several extensions of the Bogoliubov approximation have been
proposed in order to accurately reproduce the condensate fraction
and the collective mode frequencies observed in experiments. These
extensions are based on the Hartree-Fock-Bogoliubov (HFB) approximation
\cite{rf:huse}, which is closely related to the Girardeau-Arnowitt
pair theory\cite{rf:girardeau}. The most used one is the HFB-Shohno
approximation \cite{rf:shohno} (also known as HFB-Popov \cite{rf:griffin}).
However, in this approximation, the condensate-condensate correlation
is removed by hand to ensure a gapless excitation spectrum. Several
authors have suggested an improved HFB approximation by introducing
that correlation by hand so that a gapless excitation spectrum is
preserved \cite{rf:proukakis,rf:hutchinson}. More recently, Yukalov
\textit{et al.}~\cite{rf:yukalov,rf:yukalov2} proposed a treatment
of the HFB approximation in the context of representative statistical
ensembles which preserves both the correlation and absence of a gap,
and aims at describing the strongly-interacting regime.

Another line of approach at zero temperature is the variational method,
which consists in minimizing the energy within a given subspace of
the Hilbert space. The accuracy and complexity of the method depend
on the choice of the subspace. Recently, the Bijl-Dingle-Jastrow form
\cite{rf:bijl,rf:jastrow} was proposed as an interesting variational
ansatz for bosons in a lattice \cite{rf:capello}.

The purpose of this paper is to examine the {}``performance\char`\"{}
of these theories in the case of a 1D lattice system. An advantage
of lattice systems is that they eliminate issues and ambiguities associated
with ultraviolet divergences found in continuous systems with contact
interactions. We choose a 1D system because low dimensionality increases
the effects of quantum fluctuations and therefore the possible differences
between the theories. Moreover, for 1D systems, we can obtain quasi-exact
numerical solutions employing the time-evolving block decimation (TEBD)
method~\cite{rf:vidal1}, which can be used as a reference. To put
some perspective about the computational effort required by these
methods, we should note that Bogoliubov-related calculations typically
take several seconds, variational Bijl-Dingle-Jastrow calculations
several minutes, and TEBD calculations several days%
\footnote{These durations are those which were necessary for the calculations
presented in this paper and are mentioned to give some idea to the
reader, although it should be noted that they strongly depend on the
computional environment and required accuracy.%
}.

The paper is organised as follows: we first explain the TEBD method
in Section~\ref{sec:TEBD}. Especially, the explanation is focused
on the application of the TEBD method to systems with a periodic boundary
condition. The extended Bogoliubov methods (Bogoliubov, HFB, HFB-Shohno,
improved-HFB, and HFB-Yukalov) are detailed in Section~\ref{sec:HFB}.
The Bijl-Dingle-Jastrow method is presented in Section~\ref{sec:Jastrow}.
Finally, we compare all these methods numerically in Section~\ref{sec:Results}.


\section{The TEBD method\label{sec:TEBD}}

Since the main purpose of this work is to examine the performance
of several approximate approaches, accurate ground-states obtained
by a quasi-exact numerical method are necessary for a reference. The
TEBD method is a variant of the density matrix renormalization group
(DMRG) method~\cite{rf:white,rf:schollwoeck}, which is one of the
best methods available to study 1D quantum lattice systems, and provides
accurate time evolutions of many-body wave functions and ground states
via an imaginary time evolution. Although homogeneous systems with
periodic boundary conditions are better suited for the analytical
approaches that we will use later, so far the TEBD method has been
applied to systems with open~\cite{rf:vidal1} or infinite~\cite{rf:vidal2}
boundary conditions. In this section, we present a detailed explanation
on how to apply the TEBD algorithm to systems with periodic boundary
conditions.

We consider a quantum lattice system with a periodic boundary condition
composed of $M$ sites, which are labeled by index $l$, $l\in\{1,\ldots,M\}$.
We assume that the Hamiltonian consists of only on-site and nearest-neighbor
terms and is written as 
\begin{eqnarray}
\hat{H}=\sum_{l=1}^{M}\left(\hat{K}_{1}^{[l]}+\hat{K}_{2}^{[l,l+1]}\right),\end{eqnarray}
where $\hat{K}_{2}^{[M,M+1]}\equiv\hat{K}_{2}^{[M,1]}$, reflecting
the periodic boundary condition. In the case of the Bose-Hubbard model,
for example, $\hat{K}_{1}^{[l]}$ and $\hat{K}_{2}^{[l,l+1]}$ correspond
to the on-site interaction and the hopping. Spanning the Hilbert space
of the whole system by a product of local Hilbert spaces of dimension
$d$, a many-body wave function of the system can be expressed as
\begin{eqnarray}
|\Psi\rangle=\sum_{j_{1},j_{2},\ldots,j_{M}=1}^{d}c_{j_{1},j_{2},\ldots,j_{M}}|j_{1}\rangle|j_{2}\rangle\cdots|j_{M}\rangle.\end{eqnarray}
In the TEBD algorithm, coefficients $c_{j_{1},j_{2},\ldots,j_{M}}$
are decomposed in a particular matrix product form as 
\begin{eqnarray}
c_{j_{1},j_{2},\ldots,j_{M}}\!\!\! & = & \!\!\!\sum_{\alpha_{1},\ldots,\alpha_{M-1}=1}^{\chi}\!\!\!\Gamma_{\alpha_{1}}^{[1]j_{1}}\lambda_{\alpha_{1}}^{[1]}\Gamma_{\alpha_{1}\alpha_{2}}^{[2]j_{2}}\lambda_{\alpha_{2}}^{[2]}\cdots\lambda_{\alpha_{M-2}}^{[M-2]}\nonumber \\
 &  & \times\Gamma_{\alpha_{M-2}\alpha_{M-1}}^{[M-1]j_{M-1}}\lambda_{\alpha_{M-1}}^{[M-1]}\Gamma_{\alpha_{M-1}}^{[M]j_{M}}.\label{eq:tensorproduct}\end{eqnarray}
The vector $\lambda_{\alpha_{l}}^{[l]}$ represents the coefficients
of the Schmidt decomposition of $|\Psi\rangle$ with respect to the
bipartite splitting of the system into $[1,\ldots,l-1,l]:[l+1,l+2,\ldots,M]$,
\begin{eqnarray}
|\Psi\rangle=\sum_{\alpha_{l}=1}^{\chi}\lambda_{\alpha_{l}}^{[l]}|\Phi_{\alpha_{l}}^{[1,\ldots,l-1,l]}\rangle|\Phi_{\alpha_{l}}^{[l+1,l+2,\ldots,M]}\rangle,\end{eqnarray}
where the Schmidt vectors $|\Phi_{\alpha_{l}}^{[1,\ldots,l-1,l]}\rangle$
and $|\Phi_{\alpha_{l}}^{[l+1,l+2,\ldots,M]}\rangle$ are expressed
as 
\begin{eqnarray}
|\Phi_{\alpha_{l}}^{[1,\ldots,l-1,l]}\rangle\!\!\! & = & \!\!\!\sum_{j_{1},\ldots,j_{l}=1}^{d}\sum_{\alpha_{1},\ldots,\alpha_{l-1}=1}^{\chi}\!\!\!\Gamma_{\alpha_{1}}^{[1]j_{1}}\lambda_{\alpha_{1}}^{[1]}\cdots\lambda_{\alpha_{l-2}}^{[l-2]}\nonumber \\
 &  & \!\!\!\!\!\!\!\!\!\!\!\!\!\!\times\Gamma_{\alpha_{l-2}\alpha_{l-1}}^{[l-1]j_{l-1}}\lambda_{\alpha_{l-1}}^{[l-1]}\Gamma_{\alpha_{l-1}\alpha_{l}}^{[l]j_{l}}|j_{1}\rangle\cdots|j_{l-1}\rangle|j_{l}\rangle,\end{eqnarray}
and 
\begin{eqnarray}
|\Phi_{\alpha_{l}}^{[l+1,l+2,\ldots,M]}\rangle\!\!\! & = & \!\!\!\sum_{j_{l+1},\ldots,j_{M}=1}^{d}\sum_{\alpha_{l+1},\ldots,\alpha_{M}=1}^{\chi}\!\!\!\Gamma_{\alpha_{l}\alpha_{l+1}}^{[l+1]j_{l+1}}\lambda_{\alpha_{l+1}}^{[l+1]}\nonumber \\
 &  & \times\Gamma_{\alpha_{l+1}\alpha_{l+2}}^{[l+2]j_{l+2}}\lambda_{\alpha_{l+2}}^{[l+2]}\cdots\lambda_{\alpha_{M-1}}^{[M-1]}\Gamma_{\alpha_{M-1}}^{[M]j_{M}}\nonumber \\
 &  & \times|j_{l+1}\rangle|j_{l+2}\rangle\cdots|j_{M}\rangle,\end{eqnarray}
respectively. In general, the number of basis configurations $\chi$
required for convergence is of the order $d^{L/2}$ in order to express
arbitrary states~\cite{rf:vidal1}. However, since the Schmidt coefficients
$\lambda_{\alpha_{l}}^{[l]}$ decay rapidly as the index $\alpha_{l}$
increases for the ground state or low-lying excited states, the TEBD
method with relatively small number of states $\chi$ can be quasi-exact
for these states.

In order to compute the time evolution of a state, i.e. $e^{-i\hat{H}t}|\Psi\rangle$,
we split the Hamiltonian into three parts $\hat{H}_{{\rm odd}}=\sum_{1\leq m\leq M/2}(\hat{K}_{1}^{[2m-1]}+\hat{K}_{2}^{[2m-1,2m]})$,
$\hat{H}_{{\rm even}}=\sum_{1\leq m<M/2}(\hat{K}_{1}^{[2m]}+\hat{K}_{2}^{[2m,2m+1]})$,
and $\hat{H}_{{\rm edge}}=\hat{K}_{1}^{[M]}+\hat{K}_{2}^{[M,1]}$.
Subsequently, we use the second-order Suzuki-Trotter expansion to
decompose $e^{-i\hat{H}t}$ into a product of two-site operators.
When the number of lattice sites $M$ is even, 
\begin{eqnarray}
e^{-i\hat{H}\delta} & = & e^{-\frac{i\hat{H}_{{\rm odd}}\delta}{2}}e^{-i\hat{H}_{{\rm even}}\delta}e^{-i\hat{H}_{{\rm edge}}\delta}e^{-\frac{i\hat{H}_{{\rm odd}}\delta}{2}}\nonumber \\
 &  & +O(\delta^{3}),\end{eqnarray}
because $[\hat{H}_{{\rm even}},\hat{H}_{{\rm edge}}]=0$. On the other
hand, 
\begin{eqnarray}
e^{-iH\delta} & = & e^{-\frac{i\hat{H}_{{\rm odd}}\delta}{2}}e^{-\frac{i\hat{H}_{{\rm even}}\delta}{2}}e^{-i\hat{H}_{{\rm edge}}\delta}e^{-\frac{i\hat{H}_{{\rm even}}\delta}{2}}e^{-\frac{i\hat{H}_{{\rm odd}}\delta}{2}}\nonumber \\
 &  & +O(\delta^{3}),\end{eqnarray}
for odd $M$. The operators $e^{-\frac{i\hat{H}_{{\rm odd}}\delta}{2}}$
and $e^{-\frac{i\hat{H}_{{\rm even}}\delta}{2}}$ can be efficiently
applied because they act on two neighboring $\Gamma$ tensors \cite{rf:vidal1}.
However, in order to apply $e^{-i\hat{H}_{{\rm edge}}\delta}$, one
needs to deform the tensor product Eq.~(\ref{eq:tensorproduct})
to place $\Gamma^{[M]}$ next to $\Gamma^{[1]}$; this deformation
is achievable by means of the \textit{swapping techniques}~\cite{rf:shi}.

%
\begin{figure}[tb]
\includegraphics[scale=0.42]{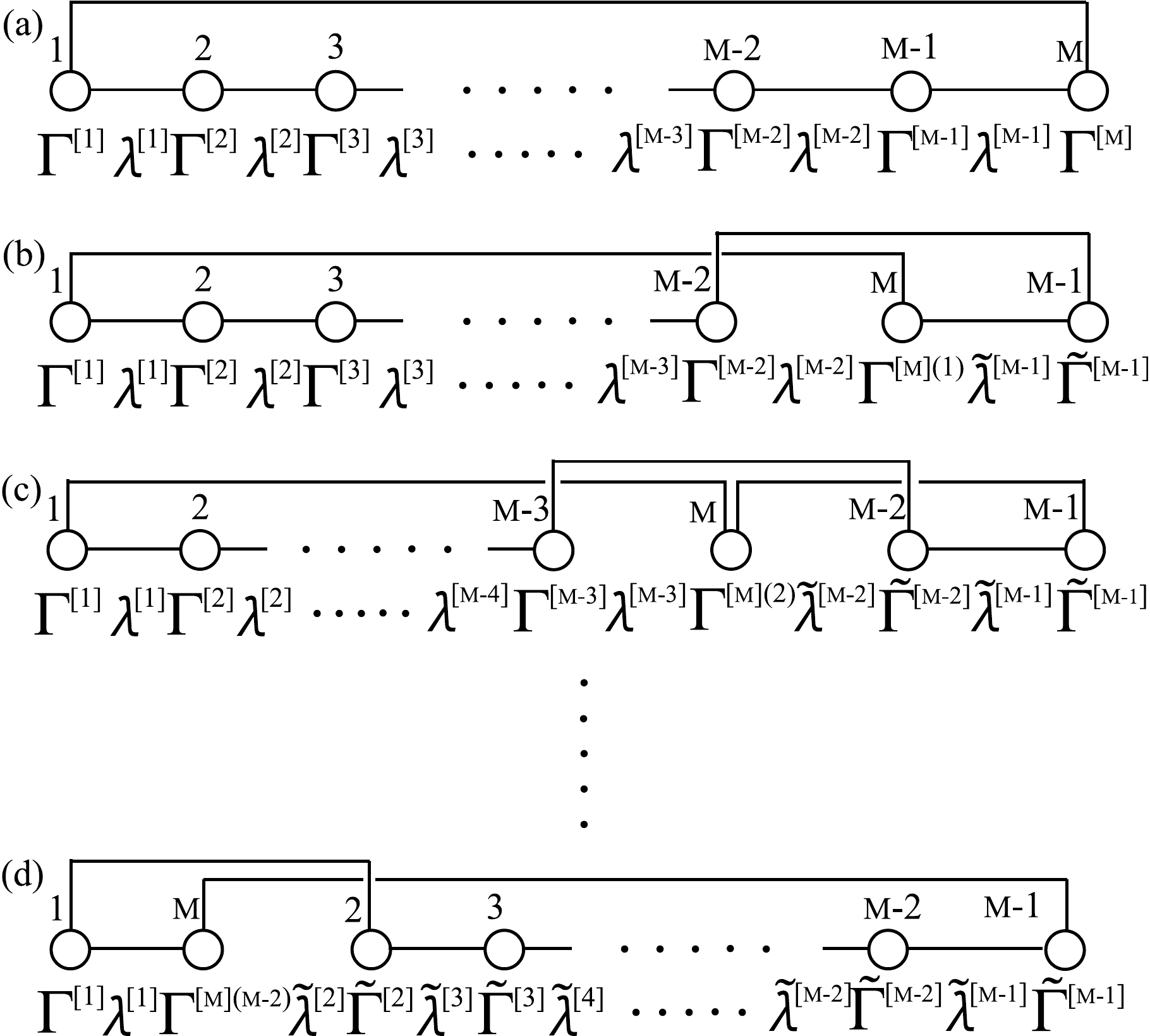} 

\caption{\label{fig:howtoPB} Schematic picture explaining how to apply $e^{-i\hat{H}_{{\rm edge}}\delta}$
to a state by using the swapping techniques. Open circles represent
lattice sites and solid lines correspond to the connections of the
lattice sites via the nearest-neighbor interactions. (a) expresses
a tensor product in the form of Eq.~(\ref{eq:tensorproduct}), In
(b), $\Gamma^{[M-1]}$ and $\Gamma^{[M]}$ are swapped and the new
tensors are expressed as $\Gamma^{[M](1)}$ and $\tilde{\Gamma}^{[M-1]}$.
In (c), subsequently, $\Gamma^{[M-2]}$ and $\Gamma^{[M](1)}$ are
swapped and the new tensors are expressed as $\Gamma^{[M](2)}$ and
$\tilde{\Gamma}^{[M-2]}$. (d) expresses a tensor product after conducting
the swapping\textcolor{red}{ $M-2$ }times. Since $\Gamma^{[1]}$
and $\Gamma^{[M](M-2)}$ are neighboring in (d), one can efficiently
apply $e^{-i\hat{H}_{{\rm edge}}\delta}$. }

\end{figure}


The swapping techniques allow us to exchange the positions of two
neighboring $\Gamma$ tensors in the tensor product and have been
successfully applied to extend the TEBD to the two-legged ladder geometry~\cite{rf:danshita}
or two-component Bose gases~\cite{rf:ludwig}. We briefly review
the swapping operation introduced in Ref.~\cite{rf:shi}, because
it is crucial for the application of the TEBD to systems with periodic
boundary conditions. For simplicity, we shall explain how to \textit{swap}
the positions of $\Gamma^{[l]}$ and $\Gamma^{[l+1]}$. Let us express
the state in a local basis that focuses on sites $l$ and $l+1$ as
\begin{eqnarray}
|\Psi\rangle\!\!\! & = & \!\!\!\sum_{j_{l},j_{l+1}=1}^{d}\sum_{\alpha_{l-1},\alpha_{l+1}=1}^{\chi}\!\!\!\Theta_{\alpha_{l-1}\alpha_{l+1}}^{j_{l}j_{l+1}}\nonumber \\
 &  & \times|\Phi_{\alpha_{l-1}}^{[1,\ldots,l-1]}\rangle|j_{l}\rangle|j_{l+1}\rangle|\Phi_{\alpha_{l+1}}^{[l+2,\ldots,M]}\rangle,\end{eqnarray}
where 
\begin{eqnarray}
\Theta_{\alpha_{l-1}\alpha_{l+1}}^{j_{l}j_{l+1}}=\sum_{\alpha_{l}=1}^{\chi}\lambda_{\alpha_{l-1}}^{[l-1]}\Gamma_{\alpha_{l-1}\alpha_{l}}^{[l]j_{l}}\lambda_{\alpha_{l}}^{[l]}\Gamma_{\alpha_{l}\alpha_{l+1}}^{[l+1]j_{l+1}}\lambda_{\alpha_{l+1}}^{[l+1]}.\label{eq:theta}\end{eqnarray}
In order to decompose $\Theta_{\alpha_{l-1}\alpha_{l+1}}^{j_{l}j_{l+1}}$
back into the tensor product form as the right-hand side of Eq.~(\ref{eq:theta}),
one has to reshape the fourth-order tensor $\Theta_{\alpha_{l-1}\alpha_{l+1}}^{j_{l}j_{l+1}}$
into a matrix $\Theta_{\{j_{l},\alpha_{l-1}\},\{j_{l+1},\alpha_{l+1}\}}$
of the dimension $d\chi\times d\chi$, where $\alpha_{l-1}$ and $\alpha_{l+1}$
are coupled with $j_{l}$ and $j_{l+1}$ respectively, and conduct
the singular value decomposition of the matrix. In the swapping process,
by coupling $\alpha_{l-1}$ with $j_{l+1}$ and $\alpha_{l+1}$ with
$j_{l}$, one reshapes the $\Theta$ tensor as $\Theta_{\alpha_{l-1}\alpha_{l+1}}^{j_{l}j_{l+1}}\rightarrow$
$\tilde{\Theta}_{\{j_{l+1},\alpha_{l-1}\},\{j_{l},\alpha_{l+1}\}}$.
The singular value decomposition of the matrix $\tilde{\Theta}_{\{j_{l+1},\alpha_{l-1}\},\{j_{l},\alpha_{l+1}\}}$
results in the swapped form of the tensor product, 
\begin{eqnarray}
\Theta_{\alpha_{l-1}\alpha_{l+1}}^{j_{l}j_{l+1}}=\sum_{\alpha_{l}=1}^{\chi}\lambda_{\alpha_{l-1}}^{[l-1]}\tilde{\Gamma}_{\alpha_{l-1}\tilde{\alpha}_{l}}^{[l+1]j_{l+1}}\tilde{\lambda}_{\tilde{\alpha}_{l}}^{[l]}\tilde{\Gamma}_{\tilde{\alpha}_{l}\alpha_{l+1}}^{[l]j_{l}}\lambda_{\alpha_{l+1}}^{[l+1]}.\end{eqnarray}
The newly obtained vector $\tilde{\lambda}_{\tilde{\alpha}_{l}}^{[l]}$
corresponds to the coefficients of the Schmidt decomposition of $|\Psi\rangle$
with respect to the bipartition of the system into $[1,\ldots,l-1,l+1]:[l,l+2,\ldots,M]$,
\begin{eqnarray}
|\Psi\rangle=\sum_{\tilde{\alpha}_{l}=1}^{\chi}\tilde{\lambda}_{\tilde{\alpha}_{l}}^{[l]}|\tilde{\Phi}_{\tilde{\alpha}_{l}}^{[1,\ldots,l-1,l+1]}\rangle|\tilde{\Phi}_{\tilde{\alpha}_{l}}^{[l,l+2,\ldots,M]}\rangle.\end{eqnarray}
where 
\begin{eqnarray}
|\tilde{\Phi}_{\tilde{\alpha}_{l}}^{[1,\ldots,l-1,l+1]}\rangle & = & \sum_{j_{l+1}=1}^{d}\sum_{\alpha_{l-1}=1}^{\chi}\lambda_{\alpha_{l-1}}^{[l-1]}\tilde{\Gamma}_{\alpha_{l-1}\tilde{\alpha}_{l}}^{[l+1]j_{l+1}}\nonumber \\
 &  & \times|\Phi_{\alpha_{l-1}}^{[1,\ldots,l-1]}\rangle|j_{l+1}\rangle,\end{eqnarray}
\begin{eqnarray}
|\tilde{\Phi}_{\tilde{\alpha}_{l}}^{[l,l+2,\ldots,M]}\rangle & = & \sum_{j_{l}=1}^{d}\sum_{\alpha_{l+1}=1}^{\chi}\!\!\!\tilde{\Gamma}_{\tilde{\alpha}_{l}\alpha_{l+1}}^{[l]j_{l}}\lambda_{\alpha_{l+1}}^{[l+1]}\nonumber \\
 &  & \times|j_{l}\rangle|\Phi_{\alpha_{l+1}}^{[l+2,\cdots,M]}\rangle.\end{eqnarray}

Once the swapping techniques are introduced, it is straightforward
to apply $e^{-i\hat{H}_{{\rm edge}}\delta}$ to a state as indicated
in Fig.~\ref{fig:howtoPB}. We first exchange positions of sites
$M-1$ and $M$ (Fig.~\ref{fig:howtoPB}(b)) and next exchange positions
of sites $M-2$ and $M$ (Fig.~\ref{fig:howtoPB}(c)). We continue
conducting the swapping until the site $M$ becomes next to the first
site as shown in Fig.~\ref{fig:howtoPB}(d). Then, $e^{-i\hat{H}_{{\rm edge}}\delta}$
can be efficiently applied. Thus, the TEBD algorithm can be applied
to systems with periodic boundary conditions by means of the swapping
techniques.

Using the {}``periodic-TEBD\char`\"{} method explained above, we
calculate the ground state of the Bose-Hubbard Hamiltonian (1) via
imaginary time evolution 
\begin{eqnarray}
|\Psi_{g}\rangle=\lim_{\tau\rightarrow\infty}\frac{e^{-\hat{H}_{{\rm bh}}\tau}|\Psi_{0}\rangle}{\parallel e^{-\hat{H}_{{\rm bh}}\tau}|\Psi_{0}\rangle\parallel},\end{eqnarray}
where $|\Psi_{0}\rangle$ is an initial state. We also use the number-conserving
version of the TEBD method, which allows substantial speed-up of the
simulations~\cite{rf:daley}. In the TEBD procedure, we choose the
maximum number of atoms per site $n_{{\rm max}}=7$ $(d=8)$ and retain
$\chi=150$ states in the adaptively selected Hilbert space.

Having obtained the ground state, we can calculate any observables
in principle by taking the average of an operator $\hat{O}$ as $\langle\hat{O}\rangle\equiv\langle\Psi_{g}|\hat{O}|\Psi_{g}\rangle$.

\section{Extended Bogoliubov methods\label{sec:HFB}}

The archetypical theory describing the weakly interacting Bose gas
near its ground state is obtained by the Bogoliubov method \cite{rf:bogoliubov}.
This method follows from the observation that the gas should be close
to a purely condensed system. One can therefore decompose the field
operator in the form $\hat{a}_{j}=\hat{c}z_{j}+\hat{\theta}_{j}$,
where $z_{j}$ is the condensate mode, and treat the noncondensate
projection $\hat{\theta}_{j}$ as a small correction. Often, it is
convenient to treat $\hat{c}$ as a number $\sqrt{N_{0}}$, where
$N_{0}$ is the number of condensed atoms, while maintaining $\langle\hat{\theta}_{j}\rangle=0$.
Although this replacement is inexact for finite-sized systems with
a fixed number of particles, it does not affect the results concerning
basic properties. The operator $\hat{\theta}_{j}$ is then regarded
as describing quantum fluctuations around the classical field $\sqrt{N_{0}}z_{j}$.

The Bogoliubov approximation corresponds to treating the fluctuations
to second order in the Hamiltonian. It results in a Hamiltonian which
is quadratic in $\hat{\theta}$, and can be diagonalised in the so-called
quasiparticle basis. The system can then be described in terms of
the noncondensate density matrix $\tilde{n}_{jk}=\langle\hat{\theta}_{j}^{\dagger}\hat{\theta}_{k}\rangle$
and noncondensate anomalous average $\tilde{m}_{jk}=\langle\hat{\theta}_{j}\hat{\theta}_{k}\rangle$.
In the ground state, for a uniform system, the condensate mode $z_{j}$
is uniform and $\tilde{n}_{jk}$ and $\tilde{m}_{jk}$ are translationally
invariant, so that we can write:\begin{align*}
z_{j} & =1/\sqrt{M}\\
\tilde{n}_{jk} & =\frac{1}{M}\sum_{q=1}^{M-1}\tilde{n}_{q}e^{i\frac{2\pi}{M}q\cdot(j-k)}\\
\tilde{m}_{jk} & =\frac{1}{M}\sum_{q=1}^{M-1}\tilde{m}_{q}e^{i\frac{2\pi}{M}q\cdot(j-k)}\end{align*}
where $\tilde{n}_{q}$ and $\tilde{m}_{q}$ are discrete Fourier transforms
of $\tilde{n}_{jk}$ and $\tilde{m}_{jk}$, and $q$ is the lattice
wave number. The $q=0$ mode (Goldstone mode) is omitted, which is
consistent with the orthogonality between the condensate and noncondensate
modes. We define the noncondensate and anomalous densities as:\begin{align}
\tilde{n} & =\frac{1}{M}\sum_{q=1}^{M-1}\tilde{n}_{q},\label{eq:n}\\
\tilde{m} & =\frac{1}{M}\sum_{q=1}\tilde{m}_{q},\label{eq:m}\end{align}
and it follows from average number conservation that\begin{equation}
n_{0}=n-\tilde{n}.\label{eq:n0}\end{equation}
where $n_{0}=N_{0}/M$ is the condensate density and $n=N/M$ is the
total density.

The quasiparticle diagonalisation of the Hamiltonian leads to the
following expressions for $\tilde{n}_{q}$ , $\tilde{m}_{q}$ in the
ground state:\begin{align}
\tilde{n}_{q} & =\frac{1}{2}\Bigg(\frac{\omega_{q}}{\sqrt{\omega_{q}^{2}-\Delta^{2}}}-1\Bigg),\label{eq:nq}\\
\tilde{m}_{q} & =-\frac{1}{2}\frac{1}{\sqrt{\omega_{q}^{2}-\Delta^{2}}},\label{eq:mq}\end{align}
 with \begin{align}
\omega_{q} & =nU+4J\sin(\frac{\pi q}{M})^{2},\label{eq:Bog1}\\
\Delta & =nU,\label{eq:Bog2}\end{align}
and the ground-state energy per site\begin{equation}
E=-2Jn+\frac{1}{2}Un^{2}+\frac{1}{2M}\sum_{q=1}^{M-1}\Bigg(\sqrt{\omega_{q}^{2}-\Delta^{2}}-\omega_{q}\Bigg).\label{eq:EnergyBogoliubov}\end{equation}
One can also show that the excitation energies as a function of the
excitation momentum are given by the quantity\begin{equation}
E_{q}=\sqrt{\omega_{q}^{2}-\Delta^{2}}.\label{eq:BogoliubovExcitation}\end{equation}

\begin{widetext}

\begin{figure}
\includegraphics[width=8cm]{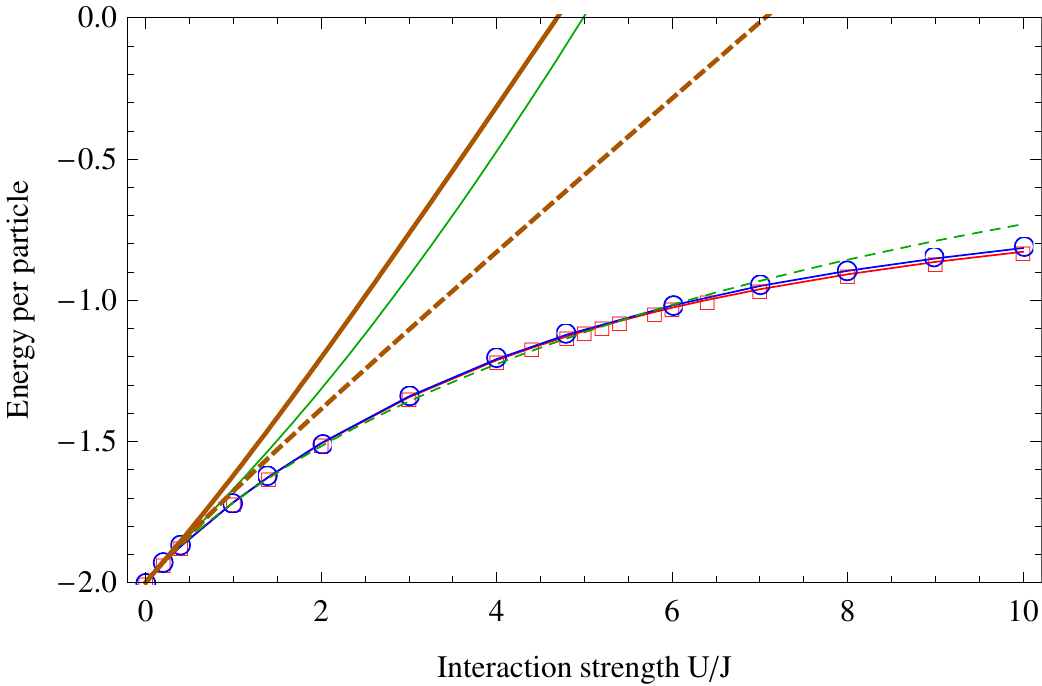}\hfill{}\includegraphics[width=8cm]{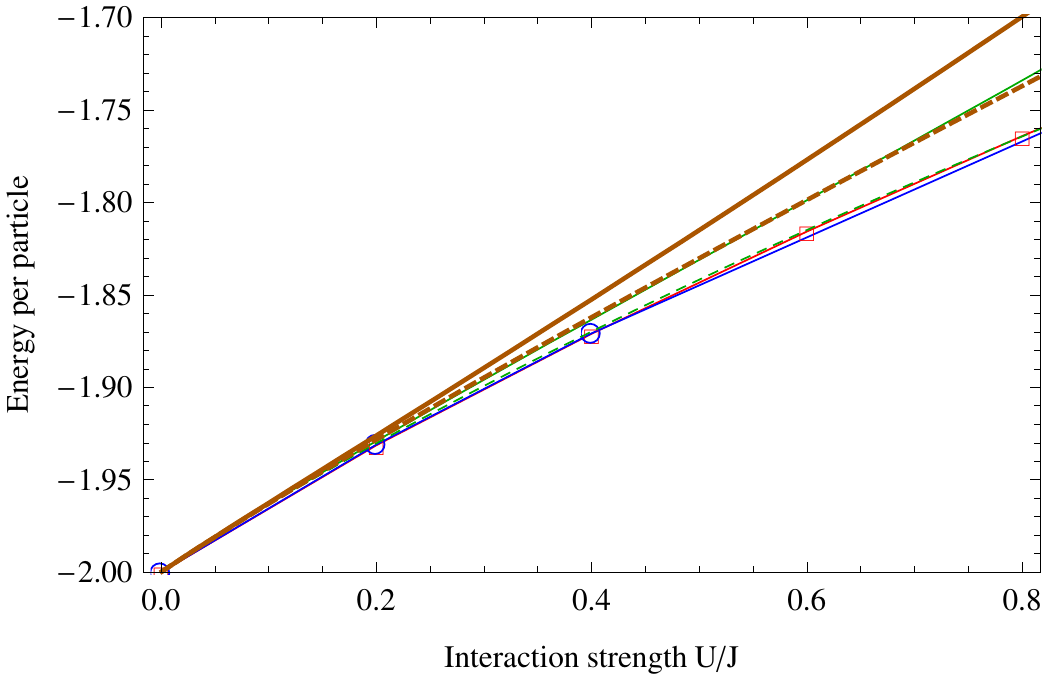}

\caption{\label{fig:Energy-for-N=00003D48}(Colour online) Energy per particle
for $N=48$, $M=60$ as a function of $U/J$. The right panel is a
close-up view of the weakly-interacting regime. The red squares correspond
to TEBD calculations, while the blue circles are obtained from the
Jastrow variational method. The thin dashed green curve corresponds
to the Bogoliubov result, the thin continuous green curve to the Popov
result, while the thick dashed orange curve is the HFB result and
the thick continuous orange curve is the improved HFB/Yukalov result.}

\end{figure}

\begin{figure}
\includegraphics[width=8cm]{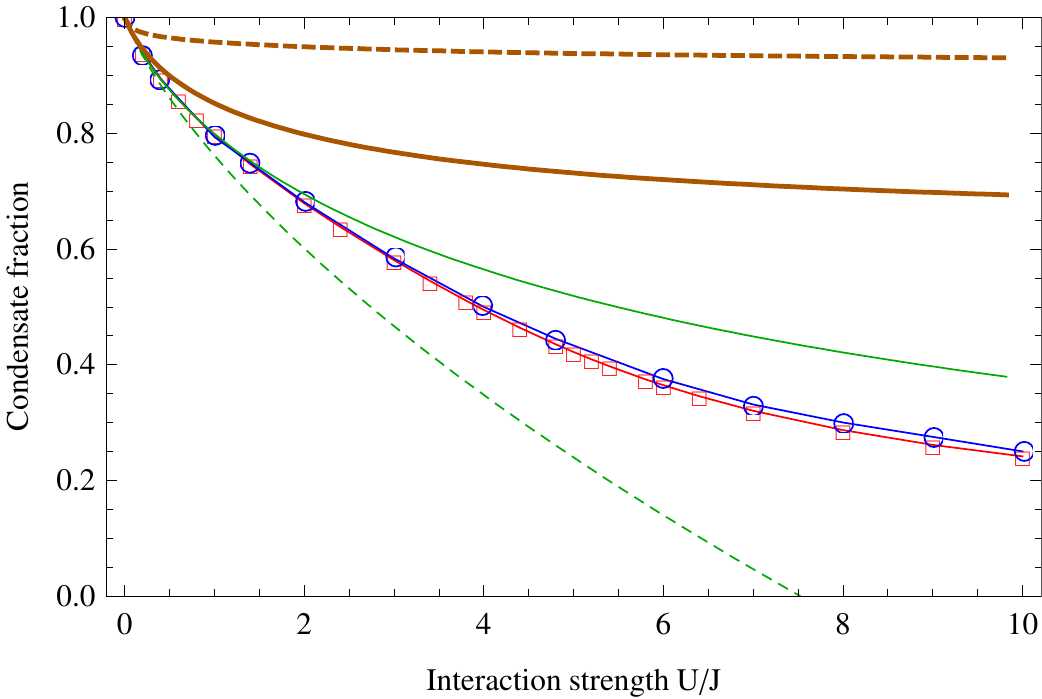}\hfill{}\includegraphics[width=8cm]{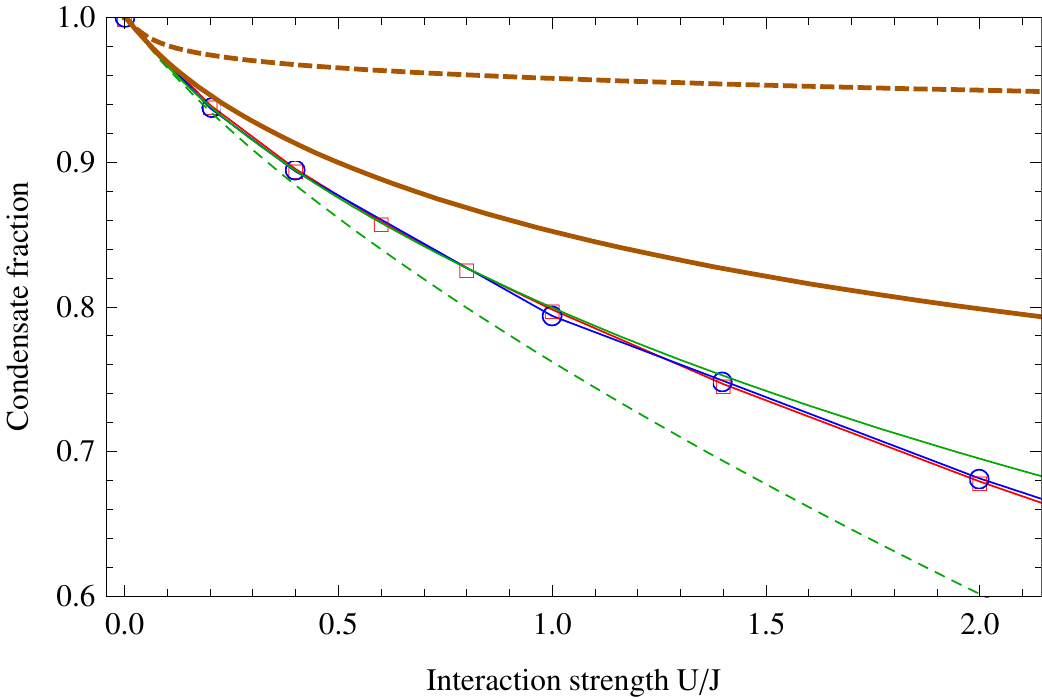}

\caption{\label{fig:Fraction-for-N=00003D48}Condensate fraction for $N=48$,
$M=60$ as a function of $U/J$. The right panel is a close-up view
of the weakly-interacting regime. The same conventions as those of
Fig.~2 are used.}

\end{figure}

\begin{figure}
\includegraphics[width=8cm]{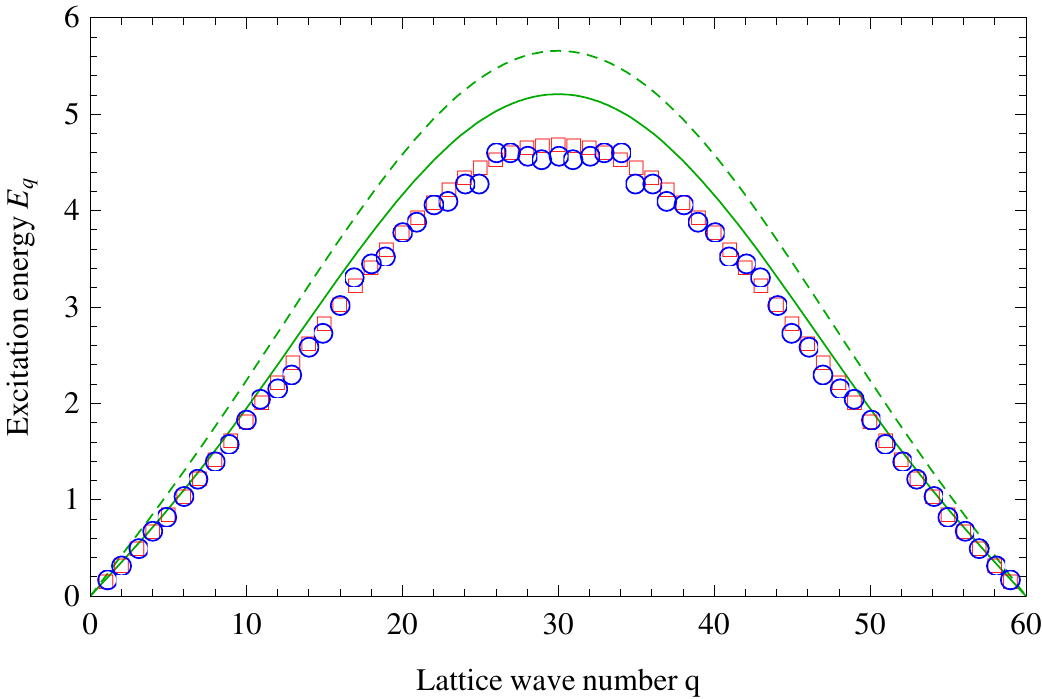}\hfill{}\includegraphics[width=8cm]{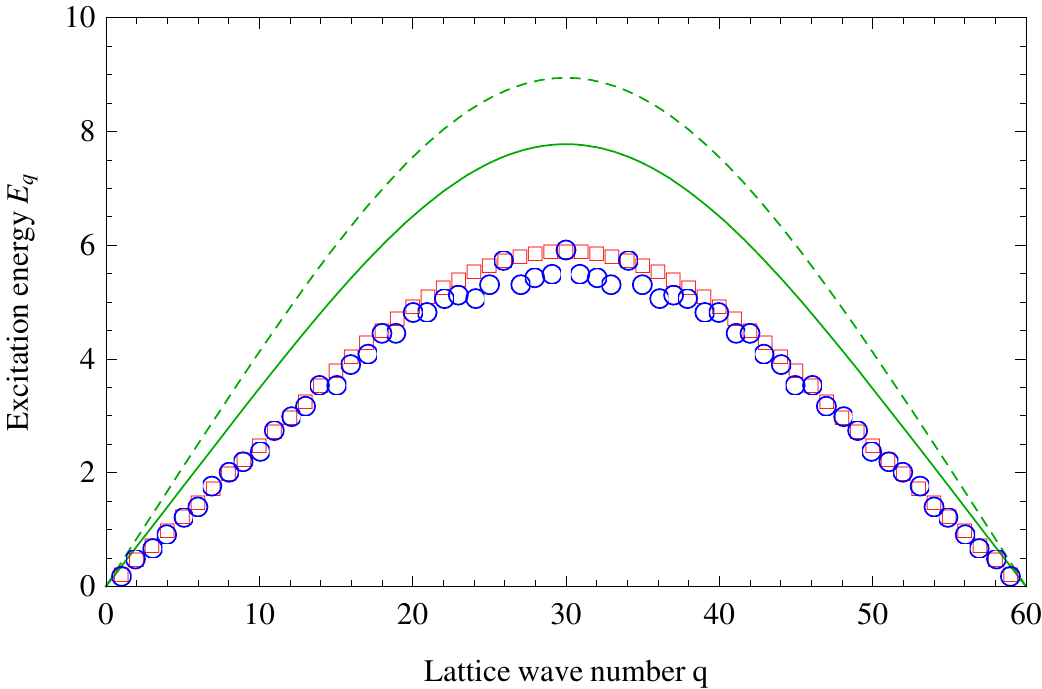}

\caption{\label{fig:Excitation-for-N=00003D48}Excitation spectrum for $N=48$,
$M=60$. The same conventions as those of Fig.~2 are used. The left
panel corresponds to $U/J=2$ and the right panel corresponds to $U/J=8$.
Note that the apparent dispersion of the blue circles are residual
fluctuations due to the stochastic nature of the variational Monte-Carlo
method.}

\end{figure}

\end{widetext}

Interactions generally tend to decrease the condensate fraction, and
as a result the Bogoliubov approximation breaks down for sufficiently
strong interactions. Especially it is unable to predict the Mott-insulator
transition in commensurate lattice systems \cite{rf:oosten}. Even
in incommensurate systems which remain superfluid, it is valid only
in the weakly-interacting limit, which can be understood as follows.
In the case of 1D Bose gases in continuum (with no lattice), the validity
of the Bogoliubov approximation is judged by the dimensionless parameter
$\gamma=l^{2}/\xi^{2}$ which expresses the competition between the
average interparticle spacing $l$ and the healing length $\xi$~\cite{rf:pitaevskii-book,rf:kinoshita}.
The healing length corresponds to the de Broglie wavelength associated
with the 1D interaction energy. When $\gamma\ll1$, the wave functions
of the particles are well overlapped with each other and the gas can
be approximately described by a macroscopic wave function (the condensate
mode) with long-range phase coherence. Therefore, the Bogoliubov approximation
is valid in this regime. As $\gamma$ increases, the overlap of the
wave functions becomes smaller and the physical quantities, such as
the ground state energy (\ref{eq:EnergyBogoliubov}) and the excitation
spectrum (\ref{eq:BogoliubovExcitation}) deviate from those of the
exact calculations based on the Bethe ansatz~\cite{rf:lieb}. In
the limit of $\gamma\gg1$, long-range coherence disappears and the
particles behave as impenetrable objects, which is referred to as
the Tonks-Girardeau gas~\cite{rf:girardeau}. In the case of the
Bose-Hubbard model, using $\xi=\hbar/(m^{\ast}c)$~\cite{rf:pitaevskii-book}
and $l=d/n$, the dimensionless parameter is expressed as $\gamma=U/(2nJ)$,
where $m^{\ast}=\hbar^{2}/(2Jd^{2})$ is the effective mass and $c=(2nJU)^{1/2}d/\hbar$
is the Bogoliubov sound speed. Thus, the Bogoliubov approximation
is valid only when $U\ll nJ$.

One way to improve the Bogoliubov approximation is to include the
higher-order perturbation terms, and make a decoupling approximation
which reduces the Hamiltonian to a quadratic form again. This is the
case when one assumes that the noncondensate particles are uncorrelated
(using Wick's decoupling scheme for the operator $\hat{\theta}$).
One then obtains the HFB approximation. The structure of the equations
is the same as above, but now\begin{align}
\omega_{q} & =(n_{0}-\tilde{m})U+4J\sin(\frac{\pi q}{M})^{2}\label{eq:HFB1}\\
\Delta & =(n_{0}+\tilde{m})U\label{eq:HFB2}\end{align}
and the energy per site is\begin{equation}
E_{HFB}=E_{Bog}+U\tilde{n}\tilde{m}+\frac{1}{2}U(\tilde{n}^{2}-\tilde{m}^{2})\label{eq:EnergyHFB}\end{equation}
where $E_{Bog}$ refers to the expression on the right-hand side of
Eq. (\ref{eq:EnergyBogoliubov}) - note that it is not the Bogoliubov
energy since $\omega_{q}$ and $\Delta$ have changed. The solution
is now implicit, and Eqs. (\ref{eq:n},\ref{eq:m},\ref{eq:n0},\ref{eq:nq},\ref{eq:mq},\ref{eq:HFB1},\ref{eq:HFB2})
have to be solved self-consistently.

While it retains the elegance of the Bogoliubov approximation, and
is supposedly applicable to more strongly-interacting cases, the HFB
approximation suffers from a notorious problem related to the existence
of a gap in the excitation spectrum \cite{rf:griffin}, which is
forbidden in the superfluid regime according to the Hugenholtz-Pines
theorem \cite{rf:FetterWalecka}. This gap can be expressed as $\lim_{q\to0}E_{q}=2U\sqrt{-\tilde{m}n_{0}}$,
and clearly comes from the anomalous average $\tilde{m}$ which is
supposed to be strictly negative, according to Eq.~(\ref{eq:mq}).
One ad hoc solution to this problem, first used by Shohno \cite{rf:shohno},
is to set $\tilde{m}$ to zero by hand in the HFB equations (\ref{eq:n},\ref{eq:n0},\ref{eq:nq},\ref{eq:HFB1},\ref{eq:HFB2},\ref{eq:EnergyHFB})
- note that this is not consistent with Eq.~(\ref{eq:mq}) and leads
to the violation of conservation laws. This solution has been popularised
as the HFB-Popov approximation in the recent literature on magnetically
trapped Bose-Einstein condensates~\cite{rf:griffin}.

However, the importance of the anomalous average near the ground state
has been stressed by some authors \cite{rf:proukakis,rf:hutchinson,rf:yukalov}
who devised new variations of the HFB approximation in order to include
that average.

In the {}``improved HFB'' approach \cite{rf:proukakis,rf:hutchinson}
it is argued that the decoupling of noncondensate particles is too
strong an assumption, resulting in noncondensate particles interacting
through a bare $T$-matrix. On the other hand, the condensate particles
interact through a full $T$-matrix, renormalised by the presence
of $\tilde{m}$, which is a measure of condensate-condensate correlations.
If one assumes that the noncondensate correlations neglected in the
HFB approximation amount to something similar to $\tilde{m}$, it
seems reasonable that the bare $T$-matrix for noncondensate particles
should be upgraded to the same renormalised $T$-matrix for condensate
particles. This can be done by hand in the HFB equations by changing
the sign in front of $\tilde{m}$ in Eq.~(\ref{eq:HFB1}). This way,
the gap disappears.

In the representative statistical ensemble approach \cite{rf:yukalov2,rf:yukalov},
it is argued that the number of condensate particles $N_{0}$ deserves
a special treatment because it appears as a new macroscopic quantity
in the system. As a result, a new chemical potential $\mu_{0}$ associated
to $N_{0}$ should be introduced. It is distinct from the chemical
potential $\tilde{\mu}$ associated to the number of noncondensate
particles $\tilde{N}=N-N_{0}$. This distinction changes Eq.~(\ref{eq:HFB1})
to\begin{equation}
\omega_{q}=(n_{0}-\tilde{m})U+(\mu_{0}-\tilde{\mu})+4J\sin(\frac{\pi q}{M})^{2}.\label{eq:omegaqHFB}\end{equation}
If one chooses $\mu_{0}-\tilde{\mu}=2\tilde{m}U$, the HFB equations
become gapless. Note that this results in the same modification as
in the {}``improved HFB'' approach. Although the two approaches
are conceptually different and do lead to different results in more
general cases (such as nonlocal interactions), they yield the same
equations in the case studied in this paper.

In both cases, the energy is obtained by formally changing the sign
of $\tilde{m}$:\begin{equation}
E_{HFBY}=E_{Bog}-U\tilde{n}\tilde{m}+\frac{1}{2}U(\tilde{n}^{2}-\tilde{m}^{2}).\label{eq:EHFBY}\end{equation}

\begin{widetext}

\begin{figure}
\includegraphics[width=8cm]{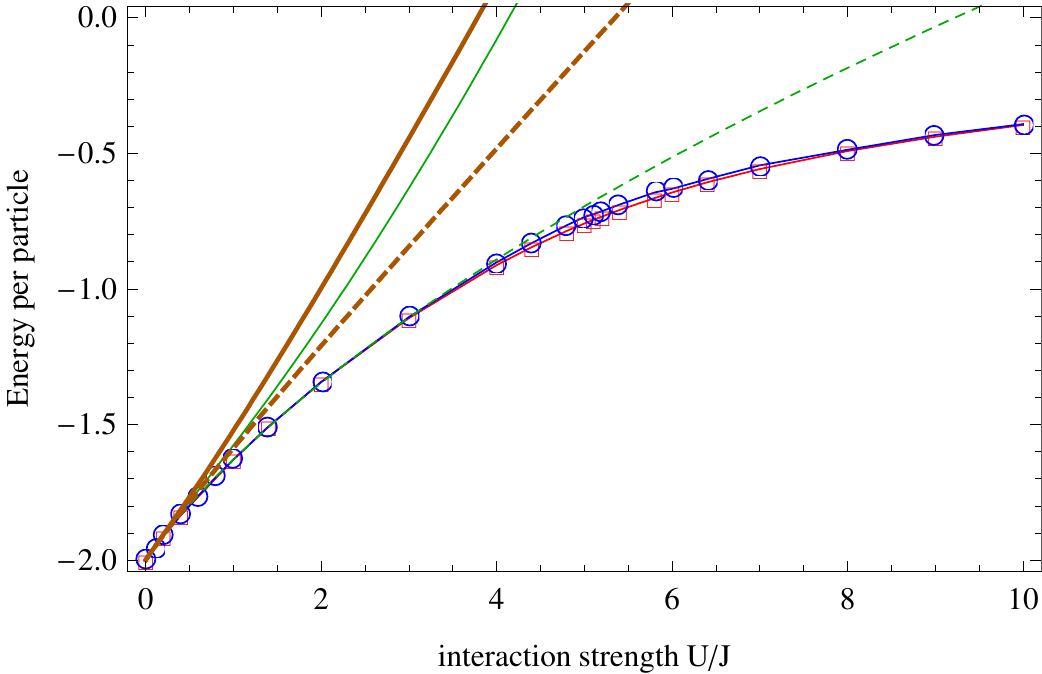}\hfill{}\includegraphics[width=8cm]{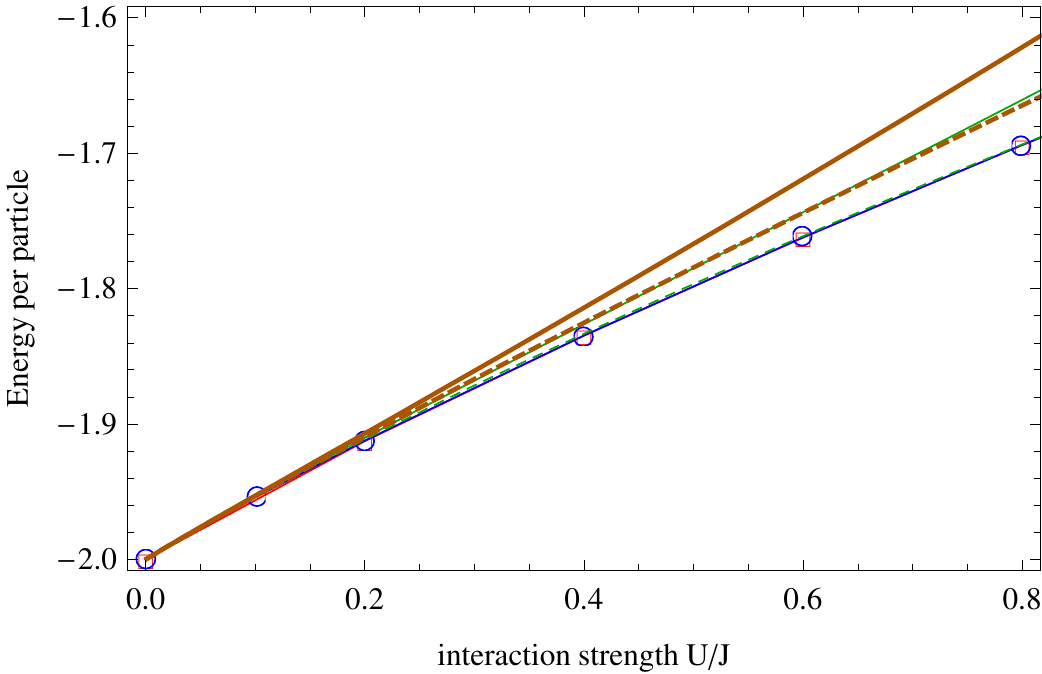}

\caption{\label{fig:Energy-for-N=00003D60}Energy for $N=60$, $M=60$ as a
function of $U/J$. The right panel is a close-up view of the weakly-interacting
regime. The same conventions as those of Fig.~2 are used.}

\end{figure}

\begin{figure}
\includegraphics[width=8cm]{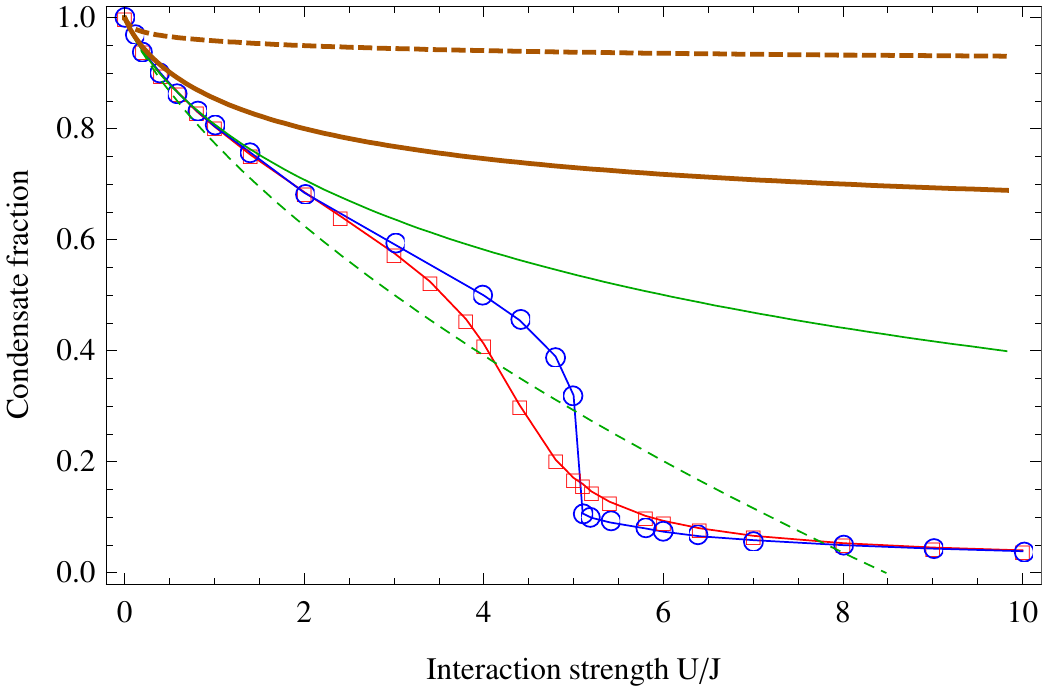}\hfill{}\includegraphics[width=8cm]{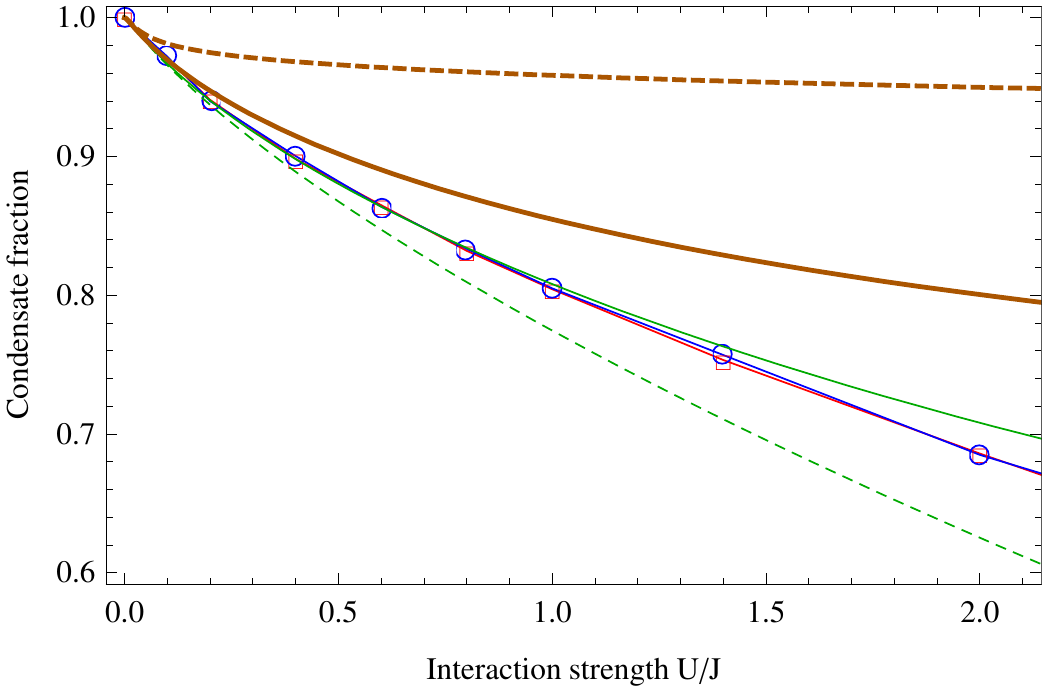}

\caption{\label{fig:Fraction-for-N=00003D60}Condensate fraction for $N=60$,
$M=60$ as a function of $U/J$. The right panel is a close-up view
of the weakly-interacting regime. The same conventions as those of
Fig.~2 are used.}

\end{figure}

\begin{figure}
\includegraphics[width=8cm]{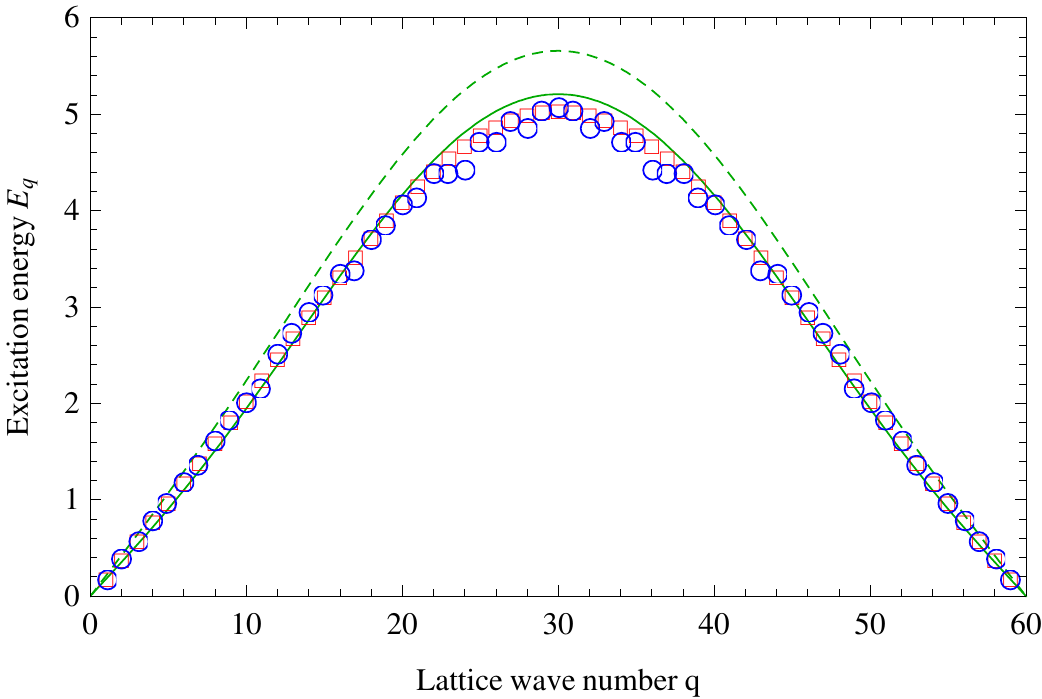}\hfill{}\includegraphics[width=8cm]{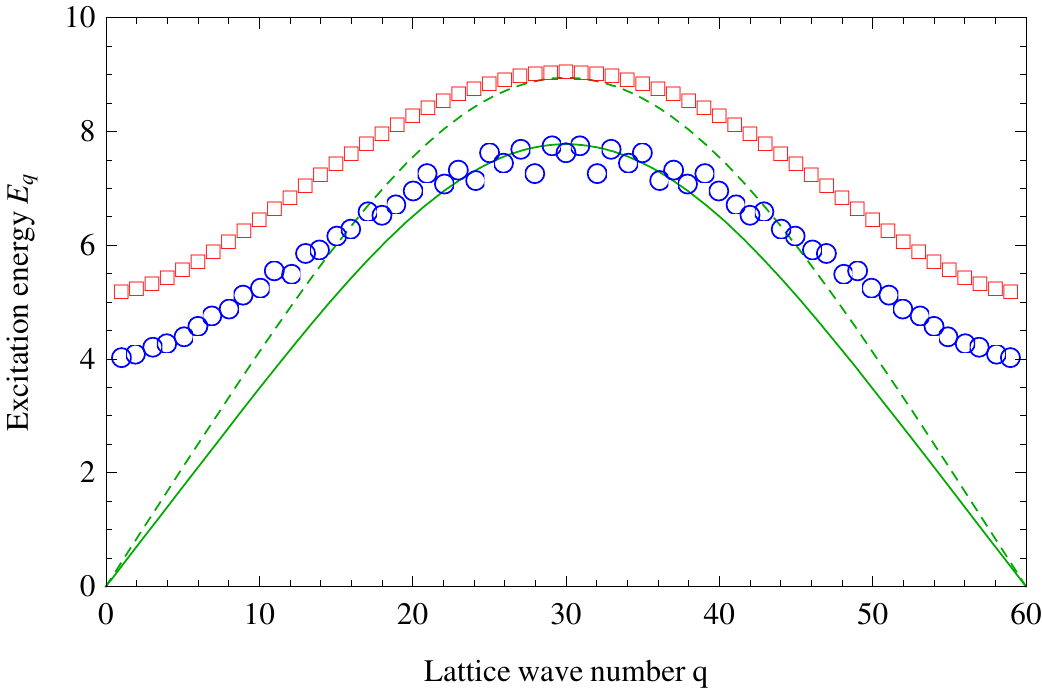}

\caption{\label{fig:Excitation-for-N=00003D60}Excitation spectrum for $N=60$,
$M=60$. The same conventions as those of Fig.~2 are used. The left
panel corresponds to $U/J=2$ and the right panel corresponds to $U/J=8$.
Note that the apparent dispersion of the blue circles are residual
fluctuations due to the stochastic nature of the variational Monte-Carlo
method.}

\end{figure}

\end{widetext}

\section{The variational Bijl-Dingle-Jastrow method\label{sec:Jastrow}}

We now consider a variational method. It consists in minimizing the
energy of the system within a subspace of the full Hilbert space.
The minimization is numerically possible if that subspace is not too
large, and leads to a good approximation of the true ground state
if it is general enough to allow the important features of that state.
As recently proposed in Ref.~\cite{rf:capello}, we choose the subspace
of wave functions having the Bijl-Dingle-Jastrow form~\cite{rf:bijl,rf:jastrow}:\begin{equation}
\Psi(x_{1}\dots x_{N})=\prod_{i<j}f(x_{i}-x_{j})\label{eq:Jastrow}\end{equation}
This is a simple form obtained by a product of some correlation function
$f$ for all pairs of bosons. The Bijl-Dingle-Jastrow form Eq. (\ref{eq:Jastrow})
is more general than the structure of the Bogoliubov ground state
(which can be reproduced from Eq. (\ref{eq:Jastrow}) in the limit
of weak interaction). Thus it can be applied successfully to more
strongly-interacting regimes. However, this method gives only the
ground-state properties and some excitation properties, unlike the
previous methods which are also applicable at finite temperature.
Another drawback is that the minimum cannot be found analytically.

To find the minimum numerically, we use the methods described in Ref.~\cite{rf:capello}.
Namely, minimization is performed by means of the power method, and
at each step a Monte-Carlo algorithm computes all the average quantities
relevant to the minimization procedure. Once the optimal $f$ is found,
any observable can be obtained by taking the average of the corresponding
operator in the quantum state $\Psi$. This average is calculated
using the Monte-Carlo algorithm.

\section{Results\label{sec:Results}}

We apply the methods described in the previous sections to solve the
Bose-Hubbard model in two cases: an incommensurate case $N=48$, $M=60$
(for which the lattice is filled to 80\%) and a commensurate case
$N=60$, $M=60$ (for which the lattice is filled to $100\%$ ). The
Superfluid-Mott insulator transition occurs only in the commensurate
case. In both cases, we look at characteristic properties of the ground
state such as its energy per particle $E$ and its condensate fraction
$n_{0}$, as well as the excitation spectrum $E_{q}$. In the extended
Bogoliubov methods, these quantities appear naturally as variables
of the problem - see Eqs.~(\ref{eq:EnergyHFB}), (\ref{eq:n0}) and
(\ref{eq:BogoliubovExcitation}). In the TEBD and Bijl-Dingle-Jastrow
methods, the energy is obtained by calculating the average $\langle\hat{H}\rangle$,
and the condensate fraction is given by $N_{{\rm 0}}/N$, where the
number of the condensate bosons $N_{{\rm 0}}$ is obtained from the
largest eigenvalue of the one-body density matrix $\langle\hat{a}_{j}^{\dagger}\hat{a}_{l}\rangle$~\cite{rf:penrose}.
A very close upper bound of the excitation spectrum $E_{k}$ is obtained
from the $f$-sum rule \cite{rf:batrouni}, which states that\[
E_{q}\lesssim K\frac{-2\sin(\frac{\pi q}{M})^{2}}{S_{q}},\]
where both the kinetic energy $K=\langle-J\sum_{i=1}^{M}(\hat{a}_{i}^{\dagger}\hat{a}_{i+1}+\hat{a}_{i+1}^{\dagger}\hat{a}_{i})\rangle$
and the structure factor $S_{q}=\frac{1}{N}\sum_{i,j}(\langle\hat{n}_{i}\hat{n}_{j}\rangle-\langle\hat{n}_{i}\rangle\langle\hat{n}_{j}\rangle)e^{iq(i-j)}$
can be calculated by quantum averages.

Results for the incommensurate case are shown in Figs.~\ref{fig:Energy-for-N=00003D48},
\ref{fig:Fraction-for-N=00003D48} and \ref{fig:Excitation-for-N=00003D48}.
One can see that there is excellent agreement between the Bijl-Dingle-Jastrow
calculations and the TEBD results for any interaction strength. On
the other hand, all the Bogoliubov methods agree with these results
only in the weakly-interacting limit $U/J\lesssim0.2$. Curiously
enough, the energy in the Bogoliubov approximation shows very good
agreement with the TEBD and Bijl-Dingle-Jastrow energy, even for strong
interactions $U/J\gg0.2$. While this might provide some analytical
insights to devise a more elaborate theory, we believe that this is
most likely a mere coincidence, since the Bogoliubov condensate fraction
is correct only for $U/J\lesssim0.2$. The same applies for the Popov
approximation which reproduces the correct condensate fraction up
to $U/J\lesssim2$ but fails to predict the average energy for $U/J\gtrsim0.2$.

Results for the commensurate case are shown in Figs.~\ref{fig:Energy-for-N=00003D60},
\ref{fig:Fraction-for-N=00003D60} and \ref{fig:Excitation-for-N=00003D60}.
One can see that a quantum phase transition for the superfluid phase
to the Mott-insulator phase occurs around $U/J=4$. Since we are dealing
with a finite-size system, the transition is smoothed. All the extended
Bogoliubov methods fail to reproduce that transition, since their
range of validity is limited to $U/J\lesssim0.2$, but again, the
Bogoliubov approximation shows a surprisingly excellent prediction
of the energy over the whole superfluid regime. As it was found in
Ref. \cite{rf:capello}, the Bijl-Dingle-Jastrow method does show
a quantum phase transition and accurately predicts the energy and
condensate fraction in the strongly-interacting limit of the Mott-insulator
regime. The excitation spectrum in Fig.~\ref{fig:Excitation-for-N=00003D60}
also shows that a gap appears in the Bijl-Dingle-Jastrow method, while
all the extended Bogoliubov spectra remain gapless (by construction).

\begin{figure}
\includegraphics[width=8cm]{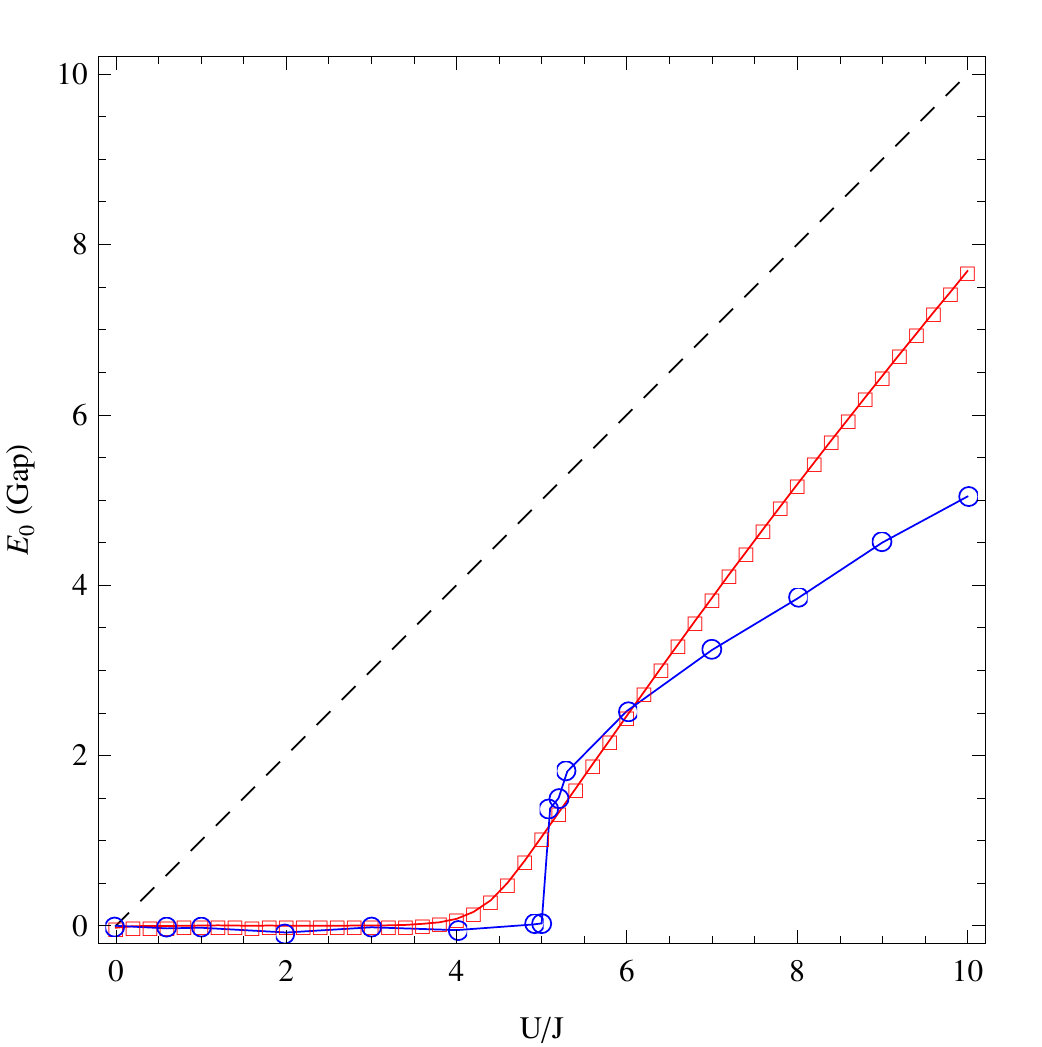}

\caption{\label{fig:gap}Gap in the excitation spectrum ($E_{q}$ for $q=0$)
as a function of $U/J$, for $N=60$, $M=60$. The red squares represent
the TEBD results, and the blue circles correspond to the variational
Bijl-Dingle-Jastrow method.}

\end{figure}

While these results seem to indicate that the Bijl-Dingle-Jastrow
ansatz is able to reproduce the superfluid-to-Mott insulator transition,
as claimed in \cite{rf:capello}, it should be noted that it does
only in a superficial way. Firstly, the transition itself is only
approximately reproduced by the Bijl-Dingle-Jastrow method, which
predicts a somewhat surprisingly sharper transition than the actual
one and at a different value of $U/J$ around 5. This fact is expected
since the system is known to be strongly correlated near the transition,
in a way which is hardly recovered by any simple ansatz. Secondly,
although the Bijl-Dingle-Jastrow ansatz tends towards the exact noninteracting
ground state when $U/J\to0$, it does not tend towards the exact Mott
insulator state when $U/J\to\infty$. Indeed, while it can reproduce
accurately the energy and condensate fraction, it fails to reproduce
the excitation gap, which even departs further from the real one as
$U/J\to\infty$, as shown in Fig.~\ref{fig:gap}. From this we conclude
that the Bijl-Dingle-Jastrow ansatz provides a partial description
a Mott-insulator, the usefulness of which is restricted to the calculation
of quantities deriving from low-order correlations. 

\section{Conclusion}

We have performed an extensive comparison of several theoretical treatments
of the Bose-Hubbard Hamiltonian ground state, describing bosonic atoms
interacting in an optical lattice. The TEBD method was generalised
to periodic boundary conditions in order to solve the problem. It
was used as a quasi-exact reference to compare with the results of
extended Bogoliubov methods and the variational Bijl-Dingle-Jastrow
ansatz.

We showed that all the methods refining the original Bogoliubov approximation
do not in fact bring any significant improvement. The validity of
all these methods is restricted to the weakly-interacting regime $U/J\lesssim0.2$.
On the other hand, the Bijl-Dingle-Jastrow ansatz proves to be a very
accurate approximation of the ground state in the superfluid phase
for any strength of the interaction, showing that particles in such
a phase are essentially correlated by pair. However, it gives only
a partial account of the ground state in the Mott insulator phase,
showing that more-than-two-particle correlations are needed in that
phase. While the structure of the Mott insulator in the strongly-interacting
regime is known, devising a theory which can account for both the
superfluid and Mott insulator phases accurately is still a challenging
problem.

\begin{acknowledgments}
I. D. is supported by a Grant-in-Aid from JSPS.
\end{acknowledgments}
\bibliographystyle{apsrev}
\bibliography{bibliopaper}

\end{document}